\documentclass[11pt]{article} 
\usepackage{mystyle-new}
\usepackage{authblk}
\usepackage[T1]{fontenc}
\usepackage{epsfig,amsmath} 
\usepackage{hyperref}
\usepackage{color}
\usepackage{graphicx}
\usepackage{orcidlink}
\definecolor{red}{rgb}{1,0,0}
\def\lesssim{\ \hbox{\raise 2pt \hbox{$<$} \kern -13pt
                     \lower 3pt \hbox{$\sim$}}\ }
\def\greatersim{\ \hbox{\raise 2pt \hbox{$>$} \kern -13pt
                     \lower 3pt \hbox{$\sim$}}\ }

\def\lsim{\mathrel{\rlap{\lower4pt\hbox{\hskip1pt$\sim$}}
    \raise1pt\hbox{$<$}}}                
\def\gsim{\mathrel{\rlap{\lower4pt\hbox{\hskip1pt$\sim$}}
    \raise1pt\hbox{$>$}}}                

\def\cascade{{\sc Cascade}}
\def\pythia{{\sc Pythia}}
\def\herwig{{\sc Herwig}}
\def\sherpa{{\sc Sherpa}}

\def\mcatnlo{{MCatNLO}}

\def\desepsf(#1 width #2){\epsfxsize=#2 \epsfbox{#1}}
\def\kt{\ensuremath{k_{\rm T}}}

\def\PZ{\ensuremath{Z}}

\def\qt{\ensuremath{q_{\rm T}}}
\def\zdyn{\ensuremath{z_{\rm dyn}}}
\def\zm{\ensuremath{z_{\rm M}}}
\def\ptll{\ensuremath{p_{\rm T}(\ell\ell)}}

\newcommand{\mdy}{\ensuremath{m_{\small\text{DY}}}}

\newcommand{\PBM}{PB}
\newcommand{\PBset}{{PB-NLO-2018}}

\newcommand{\qcdnum}{{\sc QCDnum}}

\newcommand{\MCatNLO}{{\sc MadGraph5\_aMC@NLO}}
\newcommand{\CAS}{{\mcatnlo+CAS3}}

\newcommand{\as}{\ensuremath{\alpha_s}}

\newcommand{\GeV}{\text{GeV}}
\newcommand{\TeV}{\text{TeV} }
\newcommand{\Pp}{\text{p} }

\newenvironment{tolerant}[1]{\par\tolerance=#1\relax}{ \par }
\usepackage{amsmath,bm}
\usepackage{lineno}

\usepackage{cite,./mcite}
\usepackage{tikz}
\usepackage[symbol]{footmisc}

\providecommand{\DOI}[1]{\href{http://dx.doi.org/#1}}

\begin{document}

\title{Center-of-mass energy dependence of intrinsic-\kt\ distributions obtained from Drell-Yan production} 

\author[1,2]{I.~Bubanja\thanks{itana.bubanja@cern.ch}\orcidlink{0009-0005-4364-277X}}
\affil[1]{Faculty of Science and Mathematics, University of Montenegro, Podgorica, Montenegro}
\affil[2]{Interuniversity Institute for High Energies (IIHE), Universit\'e libre de Bruxelles, Belgium} 
\author[3,4]{H.~Jung \thanks{hannes.jung@desy.de}\orcidlink{0000-0002-2964-9845}}
\affil[3]{Deutsches Elektronen-Synchrotron DESY, Germany}
\affil[4]{II. Institut f\"ur Theoretische Physik, Universit\"at Hamburg,  Hamburg, Germany}
\author[5]{A.~Lelek\thanks{aleksandra.lelek@uantwerpen.be}\orcidlink{0000-0001-5862-2775}}
\affil[5]{University of Antwerp, Belgium}
\author[1]{N.~Rai\v cevi\' c\thanks{natasar@ucg.ac.me}\orcidlink{0000-0002-2386-2290}}
\author[3]{S.~Taheri~Monfared\thanks{sara.taheri.monfared@desy.de}\orcidlink{0000-0003-2988-7859}}

\date{}
\begin{titlepage} 
\maketitle
\vspace*{-12cm}
\begin{flushright}
DESY-24-049
\end{flushright}
\vspace*{+17cm}
\end{titlepage}

\begin{abstract}
The internal motion of partons inside hadrons has been studied through its impact on
very low transverse momentum spectra of Drell-Yan (DY) pairs
created in hadron-hadron collisions. We study DY production at next-to-leading order using the Parton Branching
(PB) method which describes the evolution of transverse momentum
dependent parton distributions. The main focus is on studying the  intrinsic transverse momentum distribution  (intrinsic-\kt ) as a function of the center-of-mass energy $\sqrt s$.
While collinear parton shower Monte Carlo event generators require  intrinsic
transverse momentum distributions strongly dependent on 
$\sqrt s$, this is not the case for the PB method.
We perform a detailed study of the impact of soft parton emissions. 
We show that by requiring a minimal transverse
momentum, $q_0$, of a radiated parton, a  
dependence of the width of the intrinsic-\kt\ distribution as a function of $\sqrt{s}$ is observed. This dependence becomes stronger  with increasing $q_0$.

\end{abstract} 
\section {Introduction}

The transverse momentum distribution of Drell-Yan (DY) lepton pairs, \ptll , at large transverse momentum is well described by calculations at higher orders of the strong coupling \as , at low transverse momenta of the order of a few \GeV\ the spectrum is described by perturbative resummation, while at very low \ptll\ non-perturbative contributions become important. The resummation region can be treated in form of Transverse Momentum Dependent (TMD) parton distributions or by parton-showers in event generators like \herwig~\cite{Bellm:2015jjp,Bahr:2008pv},  \pythia~\cite{Sjostrand:2014zea,Bierlich:2022pfr} or \sherpa~\cite{Bothmann:2019yzt,Gleisberg:2008ta}. The Parton Branching method (\PBM ) \cite{Hautmann:2017fcj,Hautmann:2017xtx},  with \PBM -TMD distributions obtained from fits to inclusive HERA cross section measurements~\cite{Martinez:2018jxt},  provides an intuitive connection between parton-shower and TMD resummation.

The precise description of the  transverse momentum spectrum of DY lepton pairs at low \ptll\ at LHC energies (e.g. \cite{Bacchetta:2022awv,Bacchetta:2019sam,Martinez:2019mwt,Gauld:2021pkr,Bizon:2019zgf,Scimemi:2022ycr,Scimemi:2017etj,Echevarria:2011epo})  as well as at lower energies~\cite{BermudezMartinez:2020tys,Bacchetta:2019tcu} has been a subject for many discussions. An important role in the debate is the contribution of non-perturbative physics to the \ptll\ spectrum at very low values, at $\ptll \lesssim 1 \GeV $. In parton-shower approaches of \pythia\ and \herwig\  the intrinsic-\kt\ distribution, the transverse momentum distribution of partons at the hadron scale, plays a crucial role, and the width of this distribution is strongly dependent on the center-of-mass energy~\cite{Gieseke:2007ad, Sjostrand:2004pf}. On the contrary, predictions based on the \PBM\ approach give intrinsic-\kt\ distributions which are independent (or mildly dependent) of the center-of-mass energy and the DY mass \mdy ~\cite{Bubanja:2023nrd}. In Refs.~\cite{Bubanja:2023nrd,Mendizabal:2023mel} it is argued, that this behavior comes essentially from the treatment of soft gluons, which are included in the evolution equation, and are shown to play an important role, both for the inclusive collinear parton densities as well as for the transverse momentum distributions. These soft gluons are neglected in usual parton-shower approaches by the requirement that the emitted partons should have  transverse momenta of $\qt > q_0\simeq {\cal O}(\GeV ) $.  In Refs. \cite{Sara-PIC2023,CMS:2024aa} studies are being reported on a determination of the width of the intrinsic-\kt\ distribution to be used in parton-shower event generators \pythia\ and \herwig\  from measurements spanning  a large range of center-of-mass energies.

In this paper we give explanations of the different behavior of the intrinsic-\kt\ distributions in \PBM\ TMD and parton-shower approaches by including limitations on the value of \qt\ in calculations for TMD distributions
to mimic directly what is happening in a traditional parton-shower approach. 
It is essential to note, that no new fits for the  \PBM\ TMD have been performed, since the inclusion of a finite \qt\ cut would spoil the consistency of the evolution equation and the application of next-to-leading order (NLO) hard scattering cross sections, as shown in Ref.~\cite{Mendizabal:2023mel}. 
We will show explicitly that the inclusion of a finite  \qt\ cut leads to the observed energy dependence of the width of  the intrinsic-\kt\ distribution, stressing again the importance of a proper treatment of soft gluons for inclusive distributions.

The paper is organized as follows. In Section~\ref{sec2} we introduce the basic concept of the \PBM\ method for TMD evolution, as well as the treatment of the small transverse momentum region within this approach. We discuss how the predictions for the  transverse momentum of DY lepton pairs change with different  intrinsic-\kt\ distributions for different kinematic limits of \qt . In Section~\ref{sec3} we describe fits to DY data and evaluate the width of the intrinsic-\kt\ distributions  at different center-of-mass energies considering different limits of \qt . With Section~\ref{sec4} we conclude the paper.

\section {\label{sec2} PB TMDs and calculation of the DY cross section}

The \PBM\ method provides an elegant way to solve the DGLAP evolution equations by an iterative method simulating explicitly each individual branching that can occur during the evolution. TMD distributions are obtained with the \PBM\ method in a direct way. Essential for this method to work is the Sudakov form factor, defined at scale $\mu$:
\begin{equation}
\label{sud-def}
 \Delta_a ( \mu^2 , \mu^2_0 ) = 
\exp \left(  -  \sum_b  
\int^{\mu^2}_{\mu^2_0} 
{{d {\bf q}^{\prime 2} } 
\over {\bf q}^{\prime 2} } 
 \int_0^{\zm} dz \  z 
\ P_{ba}^{(R)}\left(\alpha_s , 
 z \right) 
\right) 
  \;\; ,   
\end{equation}
where $P_{ba}^{(R)}\left(\alpha_s ,z \right)$ are the resolvable splitting functions for splitting of parton $a$ into parton $b$, with the splitting variable $z$ being the ratio of longitudinal momenta of the involved partons. The splitting functions are explicitly given in e.g. Ref.~\cite{Hautmann:2017fcj}. The parameter \zm\ is introduced for numerical stability with $\zm = 1 - \epsilon$ with $\epsilon \to 0$. It has been shown in Ref.~\cite{Hautmann:2017fcj,Hautmann:2017xtx} that for $\epsilon$ small enough, the DGLAP limit could be reproduced and stable solutions for the inclusive as well as TMD distributions are obtained. 
The importance of the large $z$ region for inclusive and TMD distributions as well as for a parton-shower has  been discussed in detail in~\cite{Mendizabal:2023mel}. 

The integral form of  the \PBM\ evolution equation for a TMD density $ { {\cal A}}_a(x,{\bf k}, \mu^2)$  for parton $a$ at scale $\mu$  is given by:
\begin{eqnarray}
\label{evoleqforA}
   { {\cal A}}_a(x,{\bf k}, \mu^2) 
 &=&  
 \Delta_a (  \mu^2  ) \ 
 { {\cal A}}_a(x,{\bf k},\mu^2_0)  
 + \sum_b 
\int
{{d^2 {\bf q}^{\prime } } 
\over {\pi {\bf q}^{\prime 2} } }
 \ 
{
{\Delta_a (  \mu^2  )} 
 \over 
{\Delta_a (  {\bf q}^{\prime 2}  
 ) }
}
\ \Theta(\mu^2-{\bf q}^{\prime 2}) \  
\Theta({\bf q}^{\prime 2} - \mu^2_0)
 \nonumber\\ 
&\times&  
\int_x^{\zm } {{dz}\over z} \;
P_{ab}^{(R)} (\alpha_s 
,z) 
\;{ {\cal A}}_b\left({x \over z}, {\bf k}+(1-z) {\bf q}^\prime , 
{\bf q}^{\prime 2}\right)  
  \;\;  ,     
\end{eqnarray}
with $x$ being the longitudinal momentum fraction and ${\bf k}$ being the 2-dimensional vector of the transverse momentum with $\kt = |{\bf k}|$.

The intrinsic-\kt\  distribution is introduced at the starting
scale $\mu_0$ of the evolution through the distribution ${ {\cal A}}_a(x,{\bf k},\mu^2_0)$ in eq.(\ref{evoleqforA}), which is  a nonperturbative boundary condition to be determined from data. The TMD density ${ {\cal A}}_a(x,{\bf k},\mu^2_0)$ is parametrized in terms of a  collinear parton density at the starting scale and the
intrinsic-\kt\ distribution described as a  Gaussian distribution of  width $\sigma$, which is a measure of the intensity of initial intrinsic transverse motion:
\begin{equation}
\label{TMD_A0}
{\cal A}_{0,a} (x, {\bf k},\mu_0^2)   =  f_{0,a} (x,\mu_0^2)  
\cdot \exp\left(-\kt^2 / 2 \sigma^2\right) / ( 2 \pi \sigma^2) \; .
\end{equation}
The width of the Gaussian distribution $\sigma$ is related to the parameter $q_s$ via  $q_s = \sqrt{2} \sigma$. 

The \PBM\ method takes into account angular ordering by relating  the evolution scale  $|{\bf q}^\prime| = q'$  to the transverse momentum \qt : 
\begin{equation}
q' = \qt / (1-z) \; .
\label{angord}
\end{equation}
The transverse momentum of the  parton, $\bf k$, is the vectorial sum of the intrinsic transverse momentum of the initial parton and all the transverse momenta emitted in the evolution process. 
The \PBM\ evolution equation has been used to determine collinear and TMD distributions by fits to deep-inelastic measurements at HERA~\cite{Martinez:2018jxt}. Two different sets were obtained, depending of the scale choice in \as . In \PBset~Set1 the evolution scale $ q' $ was used as scale in \as , as in DGLAP evolution calculations like \qcdnum~\cite{Botje:2010ay}, leading to collinear distributions identical to the ones obtained as HERAPDF. In \PBset~Set2 the transverse momentum \qt\ was used as the scale in \as , leading to different collinear and TMD distributions. This scale choice for \as\ is motivated from angular ordering, and leads to two different regions: a perturbative region, with $\qt > q_0$, and a non-perturbative region of $\qt < q_0$. In order to avoid the divergency at the Landau pole, \as\ is frozen for $\qt < 1 $ \GeV .

The requirement of the perturbative region, $\qt > q_0$, leads directly to a restriction of $z$ as given by eq.(\ref{angord}):
\begin{equation}
  \label{zdyn} 
\zdyn = 1 - q_0/q'.
\end{equation}
Since the Sudakov form factor in eq.(\ref{sud-def})  is defined over the whole $z$ region, we can define a perturbative ($0 < z < \zdyn$) and non-perturbative ($\zdyn < z < \zm$) Sudakov form factor~\cite{Martinez:2024mou,Martinez:2024twn}:
\begin{eqnarray}
\label{eq:divided_sud}
 \Delta_a ( \mu^2 , \mu^2_0 ) & = &
\exp \left(  -  \sum_b  
\int^{\mu^2}_{\mu^2_0} 
{{d {\bf q}^{\prime 2} } 
\over {\bf q}^{\prime 2} } 
 \int_0^{\zdyn} dz \  z 
\ P_{ba}^{(R)}\left(\alpha_s , 
 z \right) 
\right) \nonumber \\
& & 
 \times \exp \left(  -  \sum_b  
\int^{\mu^2}_{\mu^2_0} 
{{d {\bf q}^{\prime 2} } 
\over {\bf q}^{\prime 2} } 
 \int_{\zdyn}^{\zm} dz \  z 
\ P_{ba}^{(R)}\left(\alpha_s , 
 z \right) 
\right) \nonumber \\
& = &  \Delta_a^{(\text{P})}\left(\mu^2,\mu_0^2,q^2_0\right)  \cdot \Delta_a^{(\text{NP})}\left(\mu^2,\mu_0^2,q_0^2\right) \; .
\end{eqnarray}
In Ref.~\cite{Bubanja:2023nrd} it was shown that $ \Delta_a^{(\text{NP})}$ plays an important role in inclusive and TMD distributions and in Ref.~\cite{Mendizabal:2023mel} it was pointed out, that neglecting $ \Delta_a^{(\text{NP})}$ can significantly affect predictions.

In parton-shower Monte Carlo event generators a minimal transverse momentum of the emitted partons is required, either in \herwig\ via the angular ordering condition and parameter 
$Q_g$~\cite{Bahr:2008pv} or in \pythia\ via $z_{max}(Q^2)$ \cite{Bierlich:2022pfr}. 
These cuts on $z$ remove completely  $\Delta_a^{(\text{NP})}$ from eq.(\ref{eq:divided_sud}).

In the following we neglect $\Delta_a^{(\text{NP})}$ (and real emissions with $z > \zdyn $)  in the TMD evolution 
to mimic the behaviour of parton-shower event generators. We do not perform new fits, but use the parameters of the starting distribution of \PBset~Set2\footnote{The \PBset~Set2 was produced with $q_0=0.01~\GeV$.} and obtain new TMD parton densities,  from {\sc updfevolv}~\cite{Jung:2024uwc}, with $q_0$ values of $1.0$ and $2.0$~\GeV\ in $\zdyn = 1 - q_0/q' $. We determine the width of the intrinsic Gauss distribution $q_s$ for the different values of $q_0$ applying the method of Ref.~\cite{Bubanja:2023nrd}, and check whether with $q_0 \sim {\cal O}(\GeV) $ we obtain an energy dependence of $q_s$  similar to the one observed in \herwig\ and \pythia .

\subsection {DY cross section at NLO}

The DY production cross section is obtained at NLO with \MCatNLO~\cite{Alwall:2014hca}, as described  and applied in Refs.~\cite{Bubanja:2023nrd,BermudezMartinez:2020tys,Martinez:2019mwt,Yang:2022qgk} using the integrated versions of the NLO parton densities \PBset~Set2. The \herwig 6 subtraction terms in MCatNLO are used since they are based on the same angular ordering conditions as the \PBM\ calculations \cite{Yang:2022qgk}. 
The \PBM\ TMD parton densities are included in the calculation via \cascade 3~\cite{Baranov:2021uol}. The simulated events (labeled as \CAS\ in the text and figures)  were passed through Rivet~\cite{Buckley:2010ar} for comparison with measurements.

\begin{figure} [htb]
\centering
\includegraphics[width=0.48\linewidth]{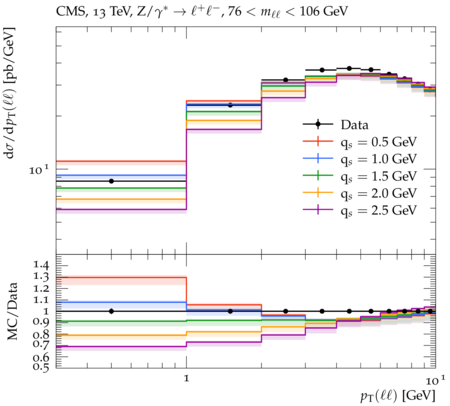} 
\includegraphics[width=0.48\linewidth]{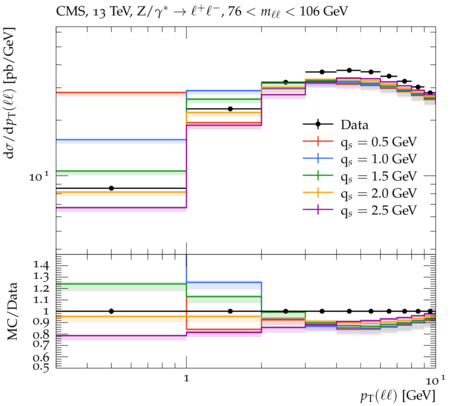}  
\caption {The DY cross section as a function of \ptll\ in the \PZ -peak region as measured by CMS~\protect\cite{CMS:2022ubq} compared to \CAS\  predictions with different
$q_s$: 0.5, 1.0, 1.5, 2.0, 2.5 \GeV , for the two values of $q_0$: $q_0 = 1$~\GeV (left) and $q_0 = 2$~\GeV (right).
The bands show the scale uncertainty.}
\label{fig:cms_q01_q02}
\end{figure}

The region of low transverse momentum of the DY lepton  pair is expected to be sensitive to the intrinsic-\kt\ distribution. We observe that this depends significantly on the region defined by the soft-gluon resolution scale \zdyn\ which is directly related to $q_0$. 
The sensitivity of the DY cross section on the intrinsic-\kt\ distribution increases with increasing cut-off $q_0$. In Fig.~\ref{fig:cms_q01_q02} we show a comparison of DY transverse momentum  distribution as measured by CMS at $\sqrt s = 13$~TeV in the \PZ\ peak region~\cite{CMS:2022ubq} with predictions obtained with the \PBM\ method with different $q_s$ values for two different scenarios of the soft-gluon resolution scale \zdyn\ (with $q_0 = 1$~GeV and $q_0 = 2$~\GeV ).

\begin{figure} [h!]
\centering
\includegraphics[width=0.48\linewidth]{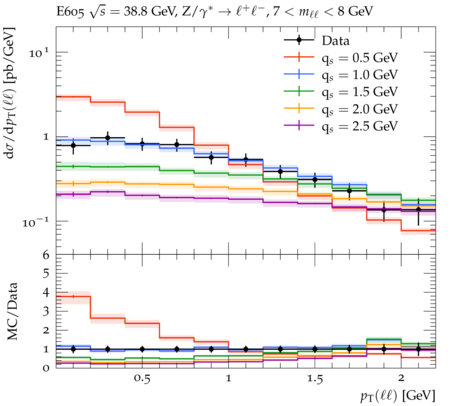} 
\includegraphics[width=0.48\linewidth]{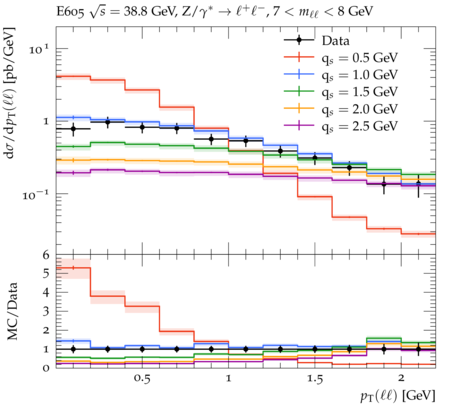}  
\caption {The DY cross section dependent on \ptll\ as measured by E605~\protect\cite{Moreno:1990sf} compared to \CAS\ predictions with different
$q_s$: 0.5, 1.0, 1.5, 2.0, 2.5 \GeV , for the two values of $q_0$: $q_0 = 1$ \GeV (left) and $q_0 = 2$~\GeV (right).
The bands show the scale uncertainty.}
\label{fig:e605_q01_q02}
\end{figure}

Using data from lower $\sqrt s$, which provide finer binning of the DY cross section at low \ptll , this sensitivity rapidly increases at very small \ptll , as shown in Fig.~\ref{fig:e605_q01_q02}, where the DY cross section measurements at $\sqrt s = 38.8$~\GeV\ obtained from E605~\cite{Moreno:1990sf} are compared to  predictions  obtained with different $q_s$ for two values of $q_0$.

\section {\label{sec3} Intrinsic-\boldmath\kt\ distribution for different \boldmath$q_0$ values}

\begin{tolerant}{8000}
The width of the intrinsic-\kt\ distribution in the \PBM\  method has been  determined  in Ref.~\cite{Bubanja:2023nrd} using  \CAS\ with the TMD set \PBset~Set2 where $q_0 = 0.01$~\GeV\ in $z_{\rm M} = 1 - q_0/q'$ . The predictions were compared  with a recent measurement  from CMS~\cite{CMS:2022ubq} on DY transverse momentum distribution in a wide range of the DY mass \mdy , obtained from {\Pp\!\!\Pp}\-collisions at $\sqrt s = 13$~TeV. A  detailed uncertainty breakdown in~\cite{Bubanja:2023nrd} in the five invariant mass bins allowed for a detailed fit. For comparison also DY measurements at lower $\sqrt{s}$ were shown.
\end{tolerant}

The width parameter $q_s$ in the TMD parton distribution was varied  and the predictions were compared to the measurements. To quantify the model agreement to the measurement, $\chi^2$ is calculated:
\begin{equation}
 \chi^2 = \sum_{i,k}(m_i - \mu_i)C_{ik}^{-1}(m_k - \mu_k)
 \end{equation}
where $m_i$ and $\mu_i$ are measurements and predictions from the $i$-th bin and $C_{ik}$ is the covariance matrix consisting of three components: a component describing the uncertainty in the measurement, the statistical (bin by bin statistical uncertainties) and scale uncertainties in the prediction.

An optimal  $q_s$ value was obtained from the minimum of the $\chi^2$ distribution with the best value for $q_s$ found to be $q_s  = 1.0\; \GeV \pm 0.08\; \GeV$.
This result was found to be consistent with $q_s$ values obtained from the measurements at lower center-of-mass energies and
only a very mild dependence of $q_s$ on $\sqrt{s}$ was observed.

\begin{figure} [h!]
\centering
\includegraphics[width=0.48\linewidth]{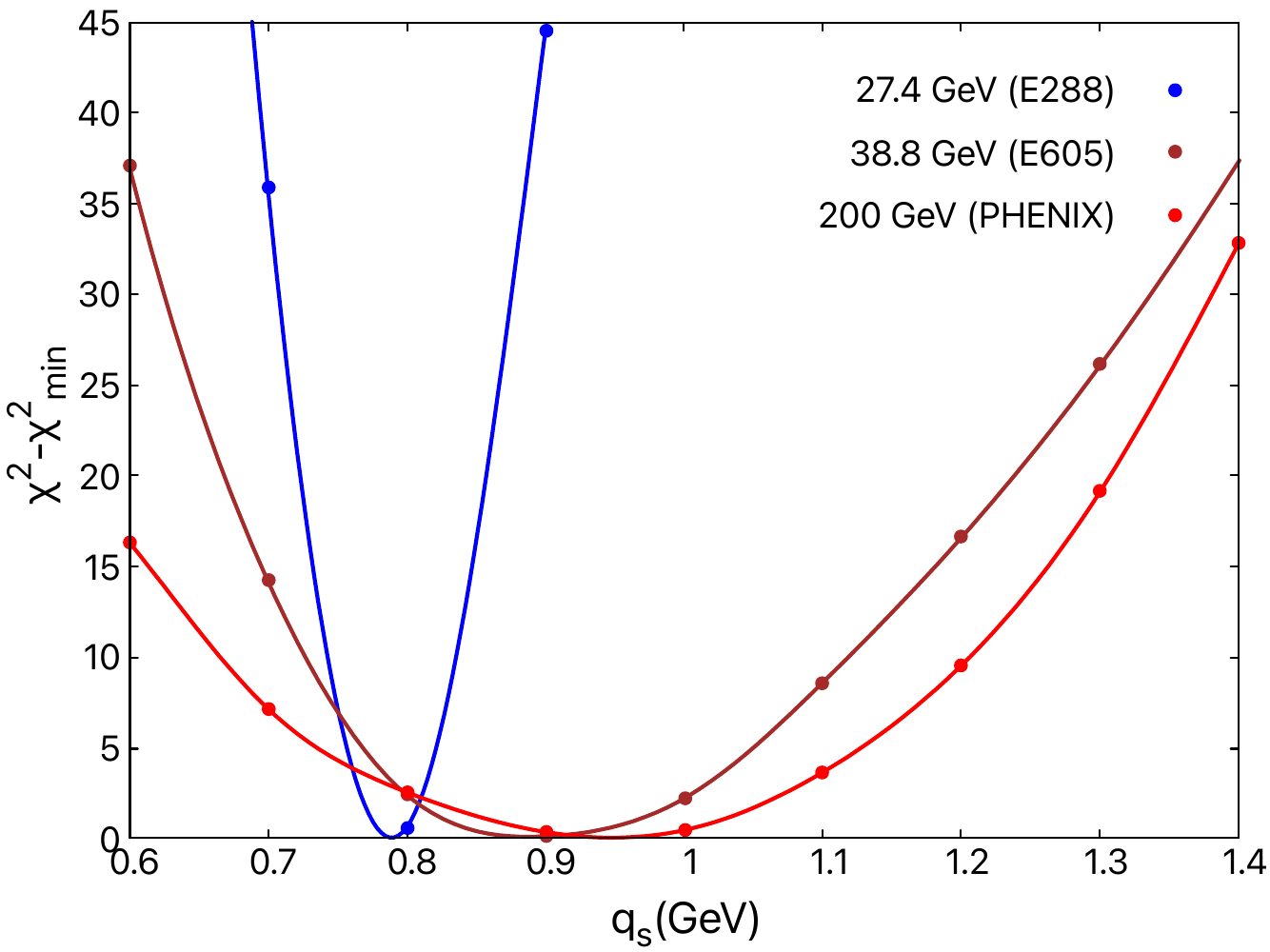} 
\includegraphics[width=0.48\linewidth]{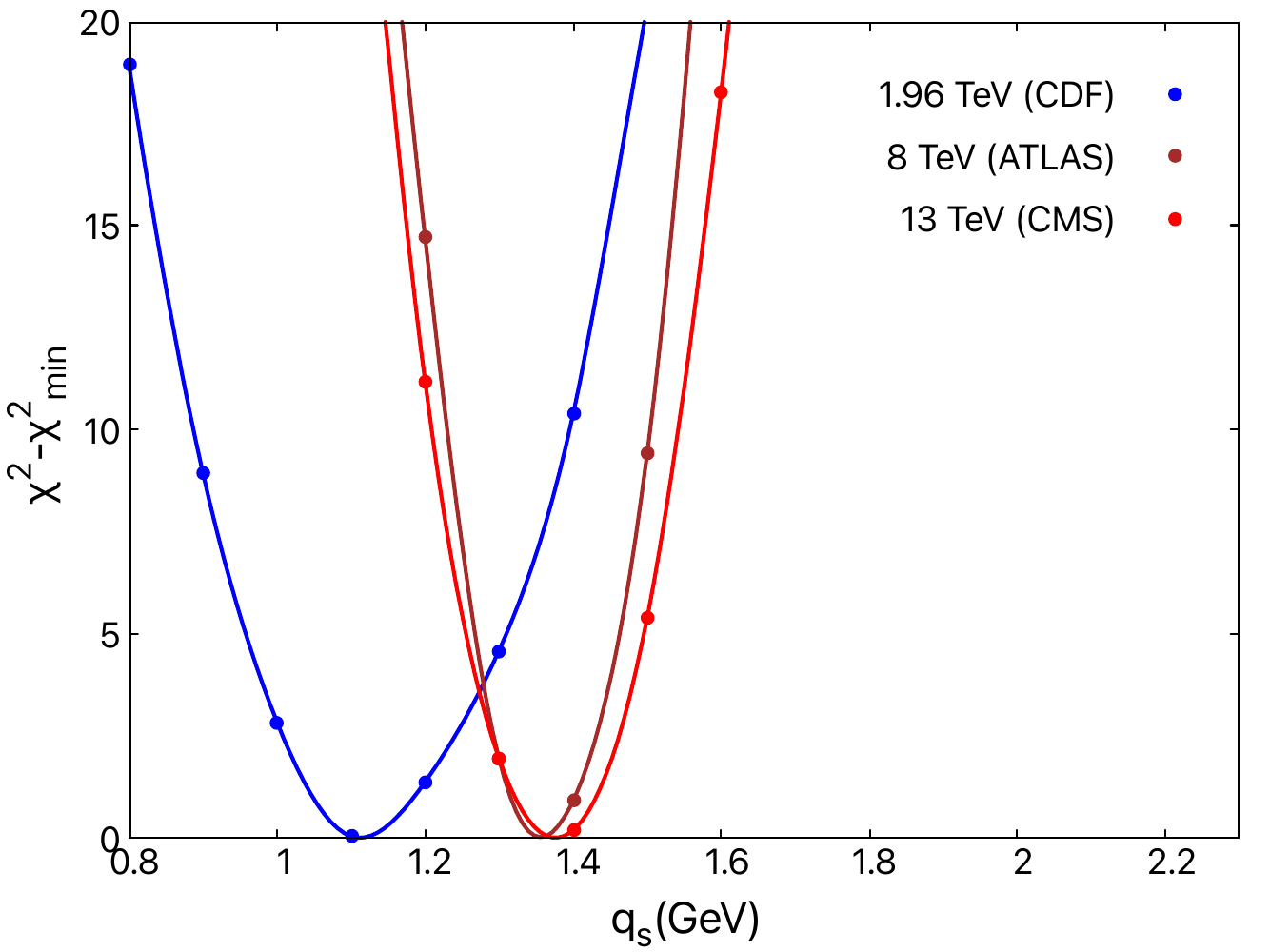}
\includegraphics[width=0.48\linewidth]{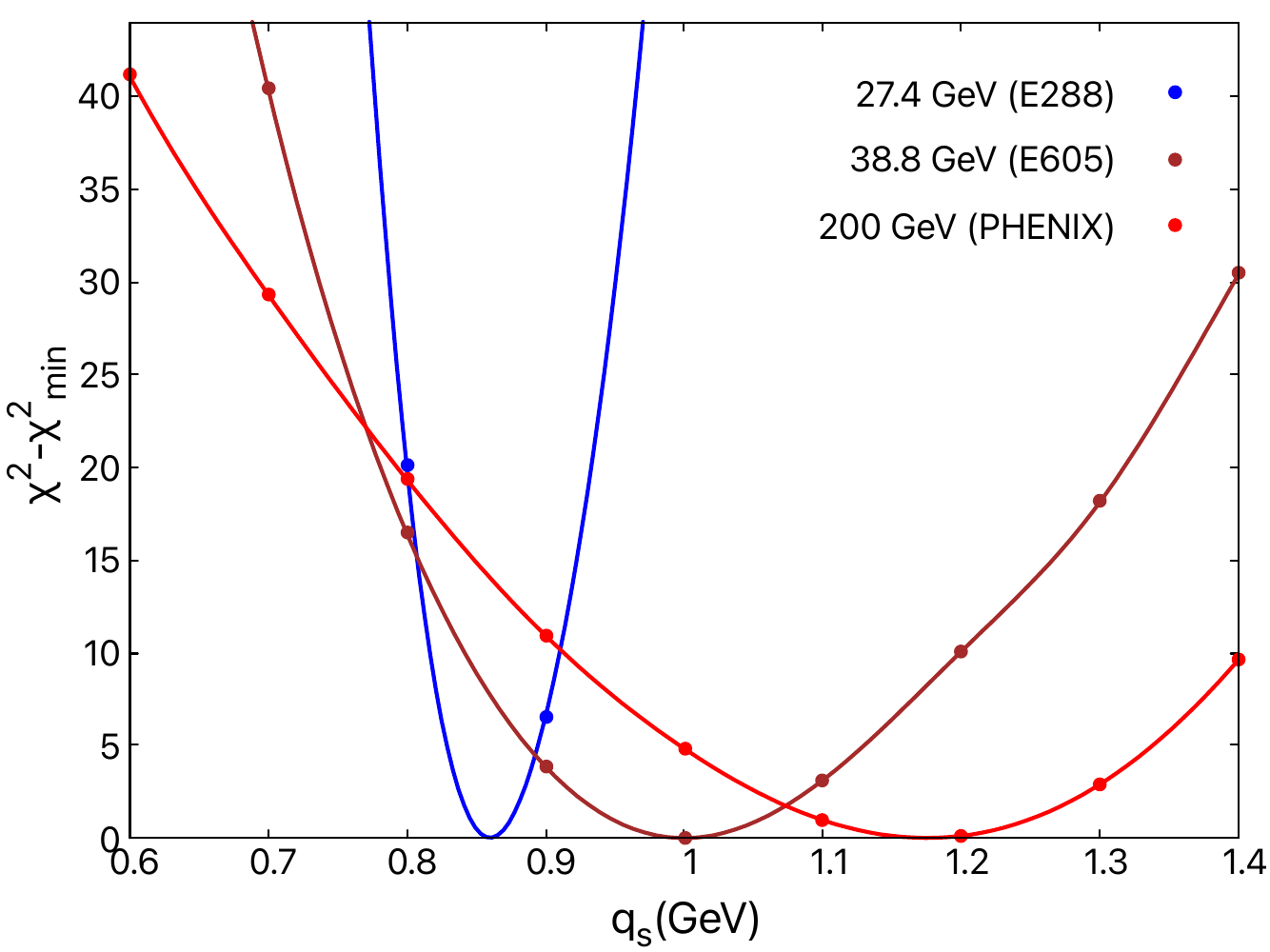} 
\includegraphics[width=0.48\linewidth]{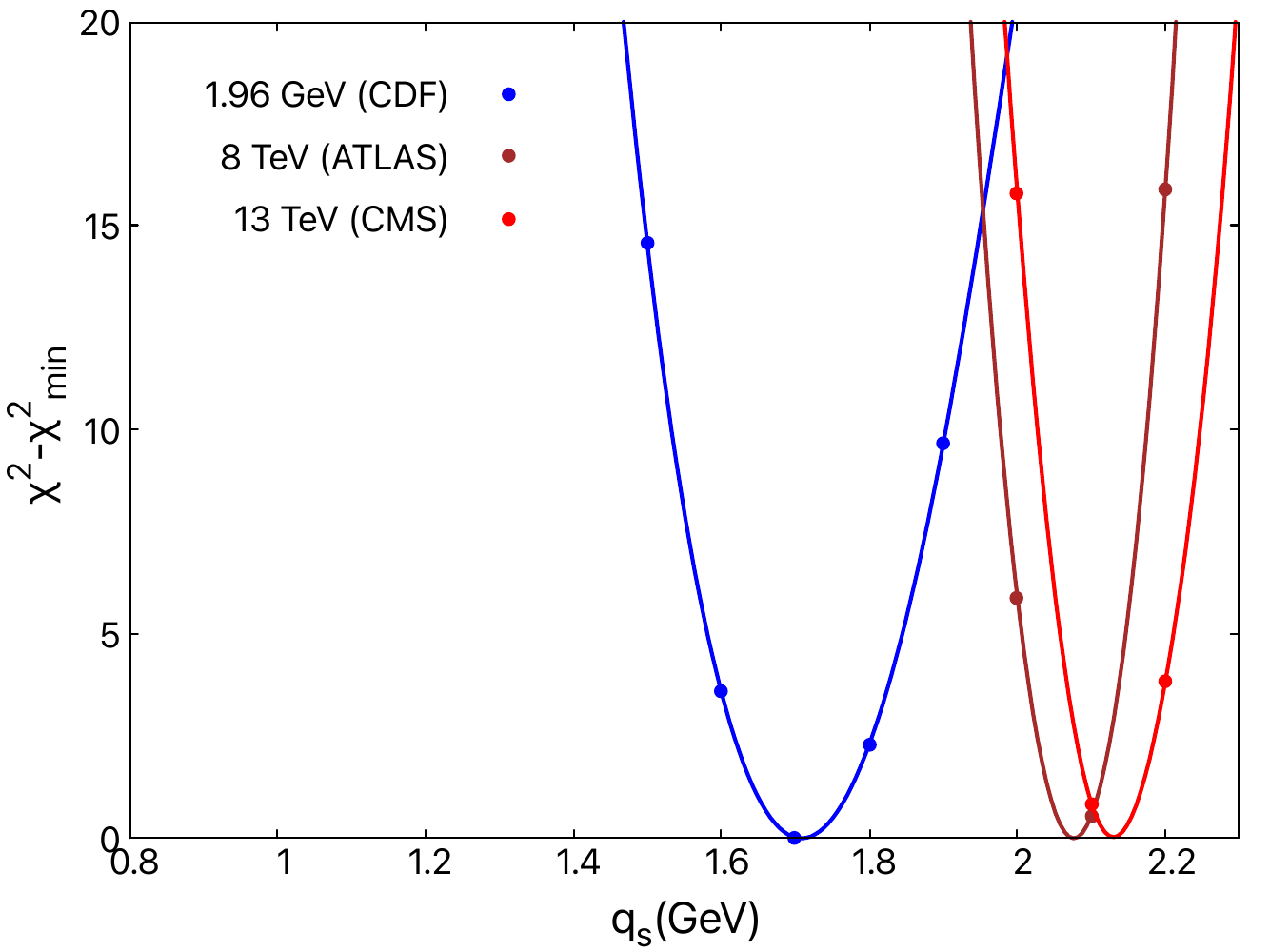}  
\caption{The $\chi^2 - \chi^2_{\rm {min}}$ distribution as a function of $q_s$ obtained from comparison of the \CAS\ prediction for $q_0 = 1$~\GeV\ (upper) and  $q_0 = 2$~GeV (lower)  with the measurements obtained at: \GeV\ energies~\protect\cite{Ito:1980ev,Moreno:1990sf,Aidala:2018ajl} (left) and \TeV\ energies~\protect\cite{CDF:2012brb,Aad:2015auj,CMS:2022ubq} (right). Each line presents a cubic spline through the points.}
\label{fig:q02_chi2}
\end{figure}

In the following we mimic parton-shower event generators by demanding a finite $q_0=1$ and $2\;\GeV $ (without performing new fits). 
With such a treatment we come as close as possible to the treatment in collinear parton-shower event generators.  
We determine $q_s$ from the experimental data given in Table.~\ref{tab:Mesurements}.
Since most of the measurements do not provide a detailed uncertainty breakdown,  we treat all the uncertainties as uncorrelated. The impact of the intrinsic-\kt\ distribution at low collision energies has been analyzed using the entire range of \ptll , while at higher center-of-mass energies, we only included bins  up to the peak region ($\ptll \simeq 8$~GeV) in the $\chi^2$ calculation.

Figure~\ref{fig:q02_chi2} shows  $\chi^2 - \chi^2_{\rm {min}}$ as a function of $q_s$ for $q_0 = 1 (2)$~GeV for low collision energies, from about 20 to 200 \GeV\ (27.4~\GeV\ from E288~\cite{Ito:1980ev}, 38.8~\GeV\ from E605~\cite{Moreno:1990sf} and 200~\GeV\ from PHENIX~\cite{Aidala:2018ajl}) as well as for high collision energies obtained at Tevatron and LHC (1.96~\TeV\ from CDF~\cite{CDF:2012brb}, 8~\TeV\ from ATLAS~\cite{Aad:2015auj} and 13~\TeV\ from CMS~\cite{CMS:2022ubq}). 
The lines shown in the figures present  $\chi^2(q_s) - \chi^2_{\rm {min}}$ with a cubic spline function interpolated through the points.

\begin{table}[h]
\centering
\begin{tabular}{c|c|c|c}
Given name   & Number of bins & CM energy [GeV] & Ref. \\
\hline
CMS & 8 & 13000 & \cite{CMS:2022ubq} \\
ATLAS   & 4 & 8000 & \cite{Aad:2015auj} \\
CDF     & 16 & 1960 & \cite{CDF:2012brb} \\
D0     & 8 & 1800 & \cite{D0:1999jba} \\
PHENIX  & 12 & 200 & \cite{Aidala:2018ajl} \\
E605    & 11 & 38.8 & \cite{Moreno:1990sf} \\
E288   & 15 & 27.4 & \cite{Ito:1980ev} \\
\end{tabular}
\caption{List of the measurements used to determine the width of the intrinsic-\protect\kt\ distribution. The number of bins in \protect\ptll\  used in the fit as well as the collision energies are given. }
\label{tab:Mesurements}
\end{table}

From the figures one can see that with increasing collision energy the minimum of $\chi^2(q_s) - \chi^2_{\rm {min}}$ shifts to higher values of $q_s$ ranging from 0.8~\GeV\ to about 1.4~\GeV\ for $q_0=1$~\GeV\ and to about 2.2~\GeV\ for $q_0=2$~\GeV .  The $\chi^2/ndf$ (with $ndf$ being the number of degrees of freedom) for all data sets is around one.
\begin{figure} [h!]
\centering
\includegraphics[width=0.8\linewidth]{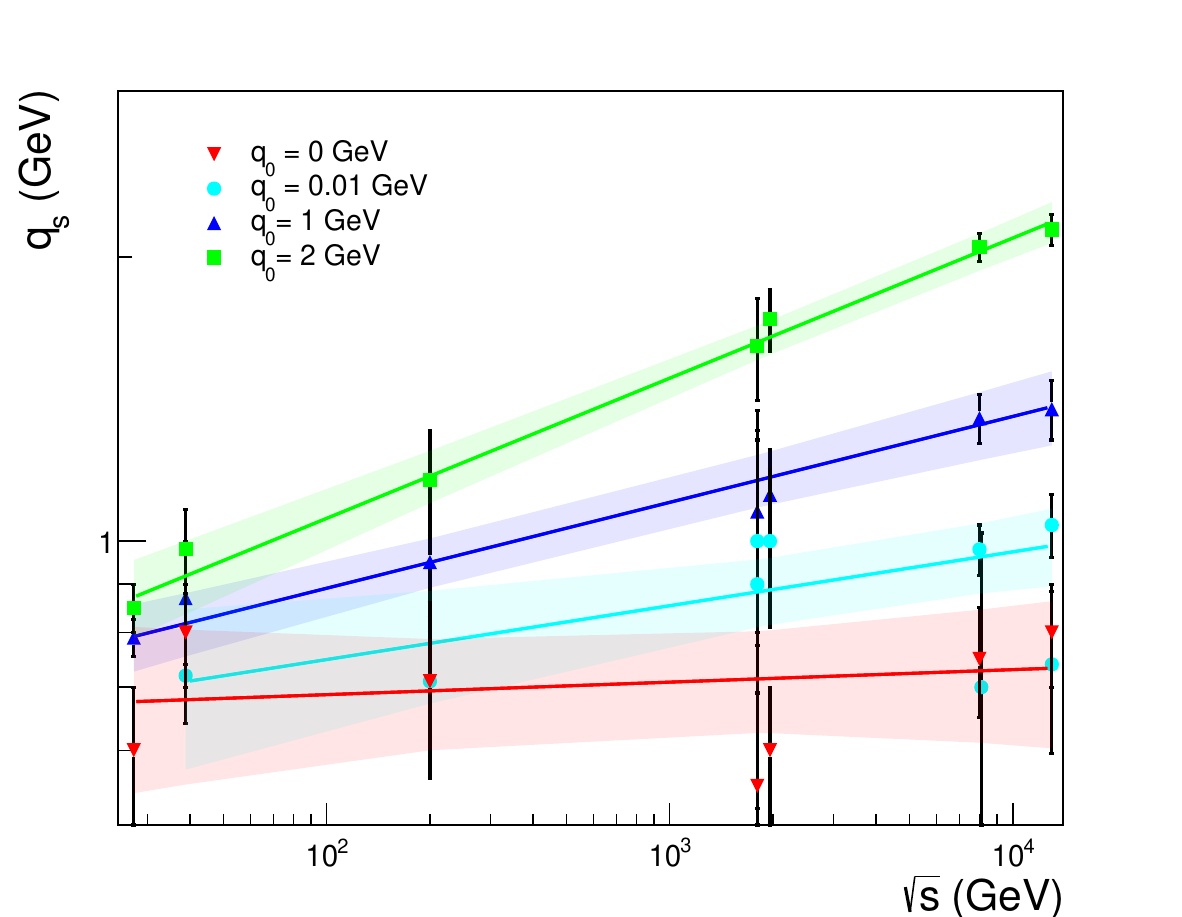}
\caption{The $q_s$ value as a function of collision energy, $\sqrt s$, obtained from the measurements presented in~\protect\cite{CMS:2022ubq,Moreno:1990sf,Ito:1980ev,Aidala:2018ajl,D0:1999jba,CDF:2012brb,Aad:2015auj} for  $q_0=0.000001$~\GeV, $q_0 = 1$~\GeV\ and  $q_0 = 2$~\GeV . Also shown are results obtained from Ref.~\protect\cite{Bubanja:2023nrd} for $q_0 = 0.01 $~\GeV .
Each line presents the linear fit of log$(q_s)$ vs log$(\sqrt s)$.}
\label{fig:energy_dependence}
\end{figure}

Summing up the results from $\chi^2(q_s)$ at different center-of-mass energies, we show  $q_s$ as a function of $\sqrt s$ in Fig.~\ref{fig:energy_dependence}. The uncertainty for each obtained $q_s$ value, which is determined as a position where $\chi^2(q_s)$ has a minimum, is estimated as a range of $q_s$ in which $\chi^2(q_s) - \chi^2_{\rm {min}} < 1$. 
The  $q_s$ dependence on the center-of-mass energy, $\sqrt s$, for the cases with  $q_0 = 1$~\GeV\ and $q_0 = 2$~\GeV\ as well as the results of Ref.~\cite{Bubanja:2023nrd} for the case $q_0 = 0.01 $~\GeV\ are shown. 
We have performed a linear fit for the $\log(q_s)-\log(\sqrt{s})$ relation. 
The uncertainty bands around the fitted lines correspond to the 95\% CL band, showing the strong anti-correlation of uncertainties between intercept and slope. 

We note that with higher $q_0$ a larger fraction of soft gluons is removed with $z < 1 - q_0/q'$  and therefore a larger contribution from intrinsic-\kt\  is needed to  accurately describe the transverse momentum spectrum in Drell-Yan processes. Consequently, higher $q_0$ values lead to an increased sensitivity to the intrinsic \kt -distribution, resulting in smaller uncertainty bands.

We observe that limiting the minimal value of transverse momentum of emitted parton at each branching by $q_0$, a dependence of $q_s$ on $\sqrt s$ is introduced.  A linear dependence of log$(q_s)$ on log$(\sqrt s)$ is observed which is confirmed by  fits with a slope  increasing with increasing~$q_0$. The result obtained in our previous study in which $q_0 =0.01$~\GeV\ is consistent with a very mild $\sqrt s$ dependence of $q_s$.  In order to confirm our findings, we calculate 
in addition predictions for $z_{\rm M} \to 1$ with $q_0=0.000001$~\GeV~\footnote{We have also performed a new fit using $q_0=0.000001$~\GeV\  to the same HERA data as used in \PBset~Set2 and found no significant differences in the collinear parton densities compared to \PBset~Set2.}.
The prediction with $q_0=0.000001$~\GeV\ clearly shows no $\sqrt s $ dependence.  We conclude, that the weak $\sqrt s$ dependence observed in Ref.~\cite{Bubanja:2023nrd} comes from $q_0 =0.01$~\GeV\  used in \PBset~Set2 and
we confirm that the dependence of the width of the intrinsic-\kt\ distribution as a function of the center-of-mass energy observed in collinear parton-shower Monte Carlo event generators comes only from the restriction of the transverse momentum of emissions in the parton-shower. No additional non-perturbative effects need to be included.

\section {\label{sec4} Conclusion}

A detailed study was performed to show the importance of soft gluon emissions in TMDs and in parton density functions in general. In this paper we confirm that the center-of-mass energy dependence of the width of the intrinsic-\kt\ distribution observed in collinear parton-shower Monte Carlo event generators comes from the treatment of soft gluons, and in particular from the non-perturbative Sudakov region, near the soft-gluon resolution boundary. 

We have studied this  effect using \PBM\  TMD distributions by imposing a cut $q_0$ restricting the $z$-integration range, in order to mimic the behavior of parton-shower event generators. In order to stay consistent with the cross section calculations, no new fits were performed, but rather the \PBM\ TMD was recalculated imposing different $q_0$ using the starting distribution of \PBset~Set2.
We have shown, that by the introduction of a finite resolution scale $q_0$ a center-of-mass energy dependent width of the intrinsic-\kt\ distribution is required by DY measurements over a wide range of  $\sqrt{s}$.
This dependence is reflected in a linear scaling of log$(q_s)$ with log$(\sqrt s)$ and the slope of this dependence increases with increasing of $q_0$.

This study emphasises the important role of soft gluons in inclusive distributions. 
The inclusion of the non-perturbative region $z \to 1$ in the evolution equation as well as in the TMD evolution is essential for the description of the low \ptll\ region in Drell-Yan production. This non-perturbative region is included  by construction  in \PBset~Set2
and this leads to a width of the intrinsic \kt -distribution independent of the collision energy $\sqrt{s}$ .

\vskip 0.5 cm 
\begin{tolerant}{8000}
\noindent 
{\bf Acknowledgments.} 
We are grateful for many fruitful discussions within the CASCADE group. 
This article is part of a national scientific project that has received funding from Montenegrin Ministry of Education, Science and Innovation. We also acknowledge funding from the European Union's Horizon 2020 research and innovation programme under grant agreement STRONG 2020 - No 824093. A. Lelek  acknowledges funding by Research Foundation-Flanders (FWO) (application number: 1272421N). S. Taheri Monfared acknowledges the support of the German Research Foundation (DFG) under grant number 467467041.

 \end{tolerant}

\bibliographystyle{mybibstyle-new.bst}

\begin{thebibliography}{10}%
\makeatletter
\providecommand{\hrefCMSnoop }[0]{\@secondoftwo}%
\makeatother
\providecommand{\doi}{\texttt{doi:}\begingroup \urlstyle{tt}\Url}

\bibitem{Bellm:2015jjp}
\hrefCMSnoop {}{J.~Bellm {et~al.}, ``{Herwig 7.0/Herwig++ 3.0 release note}'',}
  \textit{ Eur. Phys. J. C} \textbf{ 76} (2016) 196,
\href{http://www.arXiv.org/abs/1512.01178}{\texttt{arXiv:1512.01178}}.

\bibitem{Bahr:2008pv}
M.~Bahr\hrefCMSnoop {}{ {et~al.}, ``{Herwig++: physics and manual}'',} \textit{
  Eur. Phys. J. C} \textbf{ 58} (2008) 639--707,
  \href{http://www.arXiv.org/abs/0803.0883}{\texttt{arXiv:0803.0883}}.

\bibitem{Sjostrand:2014zea}
T.~Sj{\"o}strand\hrefCMSnoop {}{ {et~al.}, ``{An introduction to PYTHIA
  8.2}'',} \textit{ Comput. Phys. Commun.} \textbf{ 191} (2015) 159,
\href{http://www.arXiv.org/abs/1410.3012}{\texttt{arXiv:1410.3012}}.

\bibitem{Bierlich:2022pfr}
\hrefCMSnoop {}{C.~Bierlich {et~al.}, ``{A comprehensive guide to the physics
  and usage of PYTHIA 8.3}'',} \textit{ SciPost Phys. Codeb.} \textbf{ 2022}
  (2022) 8,
  \href{http://www.arXiv.org/abs/2203.11601}{\texttt{arXiv:2203.11601}}.

\bibitem{Bothmann:2019yzt}
\hrefCMSnoop {}{{Sherpa} Collaboration, ``{Event Generation with \mbox{SHERPA}
  2.2}'',} \textit{ SciPost Phys.} \textbf{ 7} (2019), no.~3, 034,
  \href{http://www.arXiv.org/abs/1905.09127}{\texttt{arXiv:1905.09127}}.

\bibitem{Gleisberg:2008ta}
T.~Gleisberg\hrefCMSnoop {}{ {et~al.}, ``{Event generation with SHERPA 1.1}'',}
  \textit{ JHEP} \textbf{ 0902} (2009) 007,
\href{http://www.arXiv.org/abs/0811.4622}{\texttt{arXiv:0811.4622}}.

\bibitem{Hautmann:2017fcj}
F.~Hautmann\hrefCMSnoop {}{ {et~al.}, ``{Collinear and TMD quark and gluon
  densities from Parton Branching solution of QCD evolution equations}'',}
  \textit{ JHEP} \textbf{ 01} (2018) 070,
\href{http://www.arXiv.org/abs/1708.03279}{\texttt{arXiv:1708.03279}}.

\bibitem{Hautmann:2017xtx}
F.~Hautmann\hrefCMSnoop {}{ {et~al.}, ``{Soft-gluon resolution scale in QCD
  evolution equations}'',} \textit{ Phys. Lett. B} \textbf{ 772} (2017) 446,
\href{http://www.arXiv.org/abs/1704.01757}{\texttt{arXiv:1704.01757}}.

\bibitem{Martinez:2018jxt}
A.~Bermudez~Martinez\hrefCMSnoop {}{ {et~al.}, ``{Collinear and TMD parton
  densities from fits to precision DIS measurements in the parton branching
  method}'',} \textit{ Phys. Rev. D} \textbf{ 99} (2019) 074008,
\href{http://www.arXiv.org/abs/1804.11152}{\texttt{arXiv:1804.11152}}.

\bibitem{Bacchetta:2022awv}
A.~Bacchetta\hrefCMSnoop {}{ {et~al.}, ``{Unpolarized Transverse Momentum
  Distributions from a global fit of Drell-Yan and Semi-Inclusive
  Deep-Inelastic Scattering data}'',}
  \href{http://www.arXiv.org/abs/2206.07598}{\texttt{arXiv:2206.07598}}.

\bibitem{Bacchetta:2019sam}
A.~Bacchetta\hrefCMSnoop {}{ {et~al.}, ``{Transverse-momentum-dependent parton
  distributions up to N$^{3}$LL from Drell-Yan data}'',} \textit{ JHEP}
  \textbf{ 07} (2020) 117,
  \href{http://www.arXiv.org/abs/1912.07550}{\texttt{arXiv:1912.07550}}.

\bibitem{Martinez:2019mwt}
\hrefCMSnoop {}{A.~Bermudez~Martinez {et~al.}, ``{Production of Z-bosons in the
  parton branching method}'',} \textit{ Phys. Rev. D} \textbf{ 100} (2019)
  074027,
\href{http://www.arXiv.org/abs/1906.00919}{\texttt{arXiv:1906.00919}}.

\bibitem{Gauld:2021pkr}
R.~Gauld\hrefCMSnoop {}{ {et~al.}, ``{Transverse momentum distributions in
  low-mass Drell-Yan lepton pair production at NNLO QCD}'',} \textit{ Phys.
  Lett. B} \textbf{ 829} (2022) 137111,
  \href{http://www.arXiv.org/abs/2110.15839}{\texttt{arXiv:2110.15839}}.

\bibitem{Bizon:2019zgf}
W.~Bizon\hrefCMSnoop {}{ {et~al.}, ``{The transverse momentum spectrum of weak
  gauge bosons at N ${}^3$ LL + NNLO}'',} \textit{ Eur. Phys. J. C} \textbf{
  79} (2019), no.~10, 868,
  \href{http://www.arXiv.org/abs/1905.05171}{\texttt{arXiv:1905.05171}}.

\bibitem{Scimemi:2022ycr}
\hrefCMSnoop {}{I.~Scimemi, ``{The vector boson transverse momentum
  distributions}'',} in \textit{ {56th Rencontres de Moriond~ on QCD and High
  Energy Interactions~}}.
\newblock 5, 2022.
\newblock
  \href{http://www.arXiv.org/abs/2205.05997}{\texttt{arXiv:2205.05997}}.

\bibitem{Scimemi:2017etj}
\hrefCMSnoop {}{I.~Scimemi and A.~Vladimirov, ``{Analysis of vector boson
  production within TMD factorization}'',} \textit{ Eur. Phys. J.} \textbf{
  C78} (2018), no.~2, 89,
\href{http://www.arXiv.org/abs/1706.01473}{\texttt{arXiv:1706.01473}}.

\bibitem{Echevarria:2011epo}
\hrefCMSnoop {}{M.~G. Echevarria, A.~Idilbi, and I.~Scimemi, ``{Factorization
  Theorem For Drell-Yan At Low $q_T$ And Transverse Momentum Distributions
  On-The-Light-Cone}'',} \textit{ JHEP} \textbf{ 07} (2012) 002,
  \href{http://www.arXiv.org/abs/1111.4996}{\texttt{arXiv:1111.4996}}.

\bibitem{BermudezMartinez:2020tys}
\hrefCMSnoop {}{A.~Bermudez~Martinez {et~al.}, ``{The transverse momentum
  spectrum of low mass Drell\textendash{}Yan production at next-to-leading
  order in the parton branching method}'',} \textit{ Eur. Phys. J. C} \textbf{
  80} (2020) 598,
  \href{http://www.arXiv.org/abs/2001.06488}{\texttt{arXiv:2001.06488}}.

\bibitem{Bacchetta:2019tcu}
A.~Bacchetta\hrefCMSnoop {}{ {et~al.}, ``{Difficulties in the description of
  Drell-Yan processes at moderate invariant mass and high transverse
  momentum}'',} \textit{ Phys. Rev.} \textbf{ D100} (2019), no.~1, 014018,
\href{http://www.arXiv.org/abs/1901.06916}{\texttt{arXiv:1901.06916}}.

\bibitem{Gieseke:2007ad}
\hrefCMSnoop {}{S.~Gieseke, M.~H. Seymour, and A.~Siodmok, ``{A Model of
  non-perturbative gluon emission in an initial state parton shower}'',}
  \textit{ JHEP} \textbf{ 06} (2008) 001,
\href{http://www.arXiv.org/abs/0712.1199}{\texttt{arXiv:0712.1199}}.

\bibitem{Sjostrand:2004pf}
\hrefCMSnoop {}{T.~Sj\"ostrand and P.~Skands, ``{Multiple interactions and the
  structure of beam remnants}'',} \textit{ JHEP} \textbf{ 03} (2004) 053,
\href{http://www.arXiv.org/abs/hep-ph/0402078}{\texttt{arXiv:hep-ph/0402078}}.

\bibitem{Bubanja:2023nrd}
\hrefCMSnoop {}{I.~Bubanja {et~al.}, ``{The small $k_{\textrm{T}}$region in
  Drell\textendash{}Yan production at next-to-leading order with the parton
  branching method}'',} \textit{ Eur. Phys. J. C} \textbf{ 84} (2024) 154,
  \href{http://www.arXiv.org/abs/2312.08655}{\texttt{arXiv:2312.08655}}.

\bibitem{Mendizabal:2023mel}
\hrefCMSnoop {}{M.~Mendizabal, F.~Guzman, H.~Jung, and S.~Taheri~Monfared,
  ``{On the role of soft gluons in collinear parton densities}'',} \textit{
  Eur. Phys. J. C} \textbf{ 84} (2024) 1299,
  \href{http://www.arXiv.org/abs/2309.11802}{\texttt{arXiv:2309.11802}}.

\bibitem{Sara-PIC2023}
\href {https://indico.cern.ch/event/1190468/contributions/5585083/}{\mbox{S.
  Taheri Monfared}, ``Intrinsic $k_T$ Distribution Independence in Drell-Yan
  Spectra Predictions: A Novel Insight from the Parton-Branching Method''.}
  Talk at 42nd International Symposium on Physics in Collision (PIC 2023),
  October, 2023.

\bibitem{CMS:2024aa}
\href {https://arxiv.org/pdf/2409.17770.pdf}{\mbox{CMS Collaboration},
  ``Energy-scaling behavior of intrinsic transverse momentum parameters in
  Drell-Yan simulation'',} \textit{ CMS-GEN-22-001, CERN-EP-2024-216} (09,
  2024) \href{http://www.arXiv.org/abs/2409.17770}{\texttt{arXiv:2409.17770}}.

\bibitem{Botje:2010ay}
\hrefCMSnoop {}{M.~Botje, ``{QCDNUM: fast QCD evolution and convolution}'',}
  \textit{ Comput.Phys.Commun.} \textbf{ 182} (2011) 490,
\href{http://www.arXiv.org/abs/1005.1481}{\texttt{arXiv:1005.1481}}.

\bibitem{Martinez:2024mou}
A.~B. Martinez\hrefCMSnoop {}{ {et~al.}, ``{Soft-gluon coupling and the TMD
  parton branching Sudakov form factor}'',}
  \href{http://www.arXiv.org/abs/2412.21116}{\texttt{arXiv:2412.21116}}.

\bibitem{Martinez:2024twn}
A.~\mbox{Bermudez~Martinez}\hrefCMSnoop {}{ {et~al.}, ``{The Parton Branching
  Sudakov and its relation to CSS}'',} \textit{ PoS} \textbf{ EPS-HEP2023}
  (2024) 270.

\bibitem{Jung:2024uwc}
\hrefCMSnoop {}{H.~Jung, A.~Lelek, K.~M. Figueroa, and S.~Taheri~Monfared,
  ``{The Parton Branching evolution package uPDFevolv2}'',}
  \href{http://www.arXiv.org/abs/2405.20185}{\texttt{arXiv:2405.20185}}.

\bibitem{Alwall:2014hca}
J.~Alwall\hrefCMSnoop {}{ {et~al.}, ``{The automated computation of tree-level
  and next-to-leading order differential cross sections, and their matching to
  parton shower simulations}'',} \textit{ JHEP} \textbf{ 1407} (2014) 079,
\href{http://www.arXiv.org/abs/1405.0301}{\texttt{arXiv:1405.0301}}.

\bibitem{Yang:2022qgk}
\hrefCMSnoop {}{H.~Yang {et~al.}, ``{Back-to-back azimuthal correlations in
  $\mathrm {Z} +$jet events at high transverse momentum in the TMD parton
  branching method at next-to-leading order}'',} \textit{ Eur. Phys. J. C}
  \textbf{ 82} (2022) 755,
  \href{http://www.arXiv.org/abs/2204.01528}{\texttt{arXiv:2204.01528}}.

\bibitem{Baranov:2021uol}
\hrefCMSnoop {}{S.~Baranov {et~al.}, ``{CASCADE3 A Monte Carlo event generator
  based on TMDs}'',} \textit{ Eur. Phys. J. C} \textbf{ 81} (2021) 425,
  \href{http://www.arXiv.org/abs/2101.10221}{\texttt{arXiv:2101.10221}}.

\bibitem{Buckley:2010ar}
A.~Buckley\hrefCMSnoop {}{ {et~al.}, ``{Rivet user manual}'',} \textit{ Comput.
  Phys. Commun.} \textbf{ 184} (2013) 2803,
\href{http://www.arXiv.org/abs/1003.0694}{\texttt{arXiv:1003.0694}}.

\bibitem{CMS:2022ubq}
\hrefCMSnoop {}{{CMS} Collaboration, ``{Measurement of the mass dependence of
  the transverse momentum of lepton pairs in Drell-Yan production in
  proton-proton collisions at $\sqrt{s}$ = 13 TeV}'',} \textit{ Eur. Phys. J.
  C} \textbf{ 83} (2023) 628,
  \href{http://www.arXiv.org/abs/2205.04897}{\texttt{arXiv:2205.04897}}.

\bibitem{Moreno:1990sf}
\hrefCMSnoop {}{G.~Moreno {et~al.}, ``{Dimuon production in proton - copper
  collisions at $\sqrt{s}$ = 38.8~GeV}'',} \textit{ Phys. Rev. D} \textbf{ 43}
  (1991)
2815.

\bibitem{Ito:1980ev}
\hrefCMSnoop {}{A.~S. Ito {et~al.}, ``{Measurement of the Continuum of Dimuons
  Produced in High-Energy Proton - Nucleus Collisions}'',} \textit{ Phys. Rev.}
  \textbf{ D23} (1981)
604.

\bibitem{Aidala:2018ajl}
\hrefCMSnoop {}{{PHENIX} Collaboration, ``{Measurements of $\mu\mu$ pairs from
  open heavy flavor and Drell-Yan in $p+p$ collisions at $\sqrt{s}=200$
  GeV}'',} \textit{ Phys. Rev. D} \textbf{ 99} (2019) 072003,
\href{http://www.arXiv.org/abs/1805.02448}{\texttt{arXiv:1805.02448}}.

\bibitem{CDF:2012brb}
\hrefCMSnoop {}{{CDF} Collaboration, ``{Transverse momentum cross section of
  $e^+e^-$ pairs in the $Z$-boson region from $p\bar{p}$ collisions at
  $\sqrt{s}=1.96$ TeV}'',} \textit{ Phys. Rev. D} \textbf{ 86} (2012) 052010,
  \href{http://www.arXiv.org/abs/1207.7138}{\texttt{arXiv:1207.7138}}.

\bibitem{Aad:2015auj}
\hrefCMSnoop {}{{ATLAS} Collaboration, ``{Measurement of the transverse
  momentum and $\phi ^*_{\eta }$ distributions of Drell--Yan lepton pairs in
  proton--proton collisions at $\sqrt{s}=8$ TeV with the ATLAS detector}'',}
  \textit{ Eur. Phys. J. C} \textbf{ 76} (2016) 291,
\href{http://www.arXiv.org/abs/1512.02192}{\texttt{arXiv:1512.02192}}.

\bibitem{D0:1999jba}
\hrefCMSnoop {}{{D0} Collaboration, ``{Measurement of the inclusive
  differential cross section for $Z$ bosons as a function of transverse
  momentum in $\bar{p}p$ collisions at $\sqrt{s} = 1.8$ TeV}'',} \textit{ Phys.
  Rev. D} \textbf{ 61} (2000) 032004,
  \href{http://www.arXiv.org/abs/hep-ex/9907009}{\texttt{arXiv:hep-ex/9907009}}.

\end{thebibliography}
\raggedright  
\providecommand{\href}[2]{#2}\begingroup\raggedright\endgroup

\end{document}